\documentclass[12pt]{iopart}

\usepackage{graphicx}
\usepackage{subfigure}
\usepackage{epsfig}
\usepackage{color}
\begin{document}

\title[]{Modelling Hierarchical Flocking}

\author{Yongnan Jia$^{1,2}$ \& Tamas Vicsek$^{1,3}$}

\address{$^1$ Department of Biological Physics, E\"{o}tv\"{o}s University, P\'azm\'any P\'eter s\'et\'any 1A, H-1117, Budapest, Hungary}
\address{$^2$ School of Automation and Electrical Engineering, University of Science and Technology Beijing, Beijing, P. R. China}
\address{$^3$ MTA-ELTE Statistical and Biological Physics Research Group of HAS, P\'azm\'any P\'eter s\'et\'any 1A, H-1117, Budapest, Hungary}
\ead{vicsek@hal.elte.hu}
\vspace{10pt}
\begin{indented}
\item[]October 2018
\end{indented}

\begin{abstract}
We present a general framework for modeling a wide selection of flocking scenarios under free boundary conditions. Several variants have been considered - including examples for the widely observed behavior of hierarchically interacting units. The models we have simulated correspond to classes of various realistic situations. Our primary goal was to investigate the stability of a flock in the presence of noise. Some of our findings are counter-intuitive in the first approximation, e.g., if the hierarchy is based purely on dominance (an uneven contribution of the neighbors to the decision about the direction of flight of a given individual) the flock is more prone to loose coherence due to perturbations even when a comparison with the standard egalitarian flock is made. Thus, we concentrated on building models based on leader-follower relationships. And, indeed, our findings support the concept that hierarchical organization can be very efficient in important practical cases, especially if the leader-follower interactions (corresponding to an underlying directed network of interactions) have several levels. Efficiency here is associated with remaining stable (coherent and cohesive) even in cases when collective motion is destroyed by random perturbations. The framework we present allows a the study of several further complex interactions among the members of flocking agents.
\end{abstract}
\section{Introduction}
Collective behaviour is a very important aspect through which small or large groups of organisms optimize their living \cite{Sumpter2010}. It involves collective decision making an various contexts, such as such as searching for food \cite{Petit2010}, navigating towards a distant target \cite{Petit2010, Mate2010, Vicsek2012, Kata2015} or deciding when and where to go \cite{Petit2010, Seeley2010, Anna2013}. Flocking is perhaps the most common and spectacular manifestation of collective behaviour not only in nature since recently has gained attention in the context of collective robotics as well \cite{Vasarhelyi2018, Floreano2015, Brambilla2013}. Most of the experimental and modeling approaches aimed at describing flocking by assuming egalitarian interactions among the members of a flock.

However, just as flocking is a widespread behavioural pattern of a collective, the hierarchical structure of the interactions among the members of groups is also very much common \cite{Anna2018}. Thus, starting with a trend-setting paper of Couzin et al \cite{Couzin2005} the question of leadership during flocking has attracted increasing interest. Early works assumed a two-level hierarchy while recent experimental observations involving some sophisticated  animal groups such as pigeons or primates point towards the possibility of significantly more complex internal organization principles during group motion \cite{Mate2010, Anna2013, Sueur2008, Sarova2010}. In socially highly organized groups beyond a given size (dozens or so) the roles related to leadership do not seem to be simply binary, but several levels of hierarchy can be identified. In particular the available few experimental results indicate that many co-moving groups have an internal system of interactions which can be best interpreted in terms of pairwise hierarchical levels of interactions. Perhaps the best quantitatively evaluated hierarchical group motion was carried out for pigeon flocks \cite{Mate2010, Mate2013} in which a hierarchically distributed set of interactions was demonstrated using GPS tracks and a velocity correlation-delay method \cite{Mate2010}.

On the other hand, in spite of its relevance, there have been no realistic models proposed and studied devoted to the understanding of the conditions under which hierarchical flocking can be optimal. The only vaguely related original model is in a beautiful work by Shen \cite{Shen2006}, but the assumptions are quite arbitrary and the effect of the hierarchy on the performance of the flock is not investigated.

Therefore, our goal here is to design a general framework which allows the treatment of questions related to hierarchical leader-follower interactions in a general setting, when the flock is freely moving (in the present study in two dimensions). We also discuss the various variants allowed by our approach and characterize the efficiency of the behavioral rules by determining to what level a flock is stable against external perturbations. Hierarchy is thought to be prevalent because it can be shown to result in more efficient group performance \cite{Anna2013, Anna2018}.

%further (overlapping) suggestions for the intro part:

The present work is the first one in which a flexible and plausible framework is introduced to model hierarchical flocking. It turns out that introducing pairwise interactions satisfying a realistic hierarchical dynamics is far from being trivial. After introducing the model the main aspect we investigate is the stability of a flock moving freely in a border-less two-dimensional space. We shall associate with stability the tendency of a flock not to break into parts under external perturbations.

\section{Definitions used in our model}

For an egalitarian system, every particle is believed to be the same, while for a hierarchical system, individual difference exists. Any complex multi-agent system can be classified as egalitarian or hierarchical according to the set of pairwise relationships between any two individuals. For each pair of particles in an egalitarian system, both of them have the same ability/contribution to the influence on the decision for the next time step. Meanwhile, for each pair of particles in a hierarchical system, when making a decision, they may have different level of contribution to the decision (weight) or have a directed information flow relationship, like the leader-follower mechanism (follower has no influence on the behavior of the leader).

%\textcolor[rgb]{1.00,0.00,0.00}{We assume that two particles interact if they are within a given distance. }

Graphs/networks represent a useful tool for visualizing these pairwise interaction relationships of individuals belonging to one co-moving group. Therefore, we define several matrices to describe the internal properties of the hierarchical mechanisms from different points of view, in order to characterize the differences between the egalitarian and hierarchical systems more clearly.

\subsection{Contribution matrix}

Contribution matrix $C_N=[c_{ij}]_{N\times N}(c_{ij}> 0)$ is defined to describe the contribution strength (weight) of each particle during the decision making process regarding the new preferred directions of the particles.
\begin{enumerate}
\item For an egalitarian system, every particle has the same contribution value, that is, $c_{ij}=q(q > 0)$, for $i, j=1, \cdots, N$. $q$ is a constant.
\item For a hierarchical system, not all of these $c_{ij}$ ($i, j=1, \cdots, N$) have the same positive value. For example, $c_{ij}=q_i$($q_i > 0$), for $i,j=1,\cdots,N$, or $c_{ij}$ satisfies some probability distribution(such as log-normal), for $i,j=1,\cdots,N$. We name this kind of hierarchy as \textit{contribution driven hierarchical system}.
\end{enumerate}

\subsection{Dominance Matrix}

The dominance matrix $B_N=[b_{ij}]_{N\times N}$ is defined to describe the direction of information flow between each pair individuals.

\begin{enumerate}
\item For each pair of particles $i$ and $j$ in an egalitarian system, their behaviors can be influenced by each other, that is to say, the information between each pair of particles is transmitted bidirectionally (corresponding to an undirected graph). Thus, in an egalitarian system for each pairwise interaction $b_{ij}=b_{ji}=1$, $\forall i,j=1,\cdots,N$.
\item If the information flow of paired particles is directional, that is, only one particle can obtain the other particle's information (directed graph), then we have a dominance driven mechanism, which is another kind of hierarchy. For paired particle $i$ and $j$, if $i$ is led by $j$, then we have $b_{ij}=1$ and $b_{ji}=0$, $i,j=1,\cdots,N$.
\end{enumerate}

Leader-follower mechanism is a typical kind of dominance relation in hierarchical organizations. For each pair of particles, leader particle does not take into account the influence of the follower, but the follower considers the behavior of the leader particle when makes decision on its behavior at the next step. This feature is represented by the fact that the matrix $B_N$ is not symmetric. Without loss of generality, we number these particles according to the level of dominance from $1$ to $N$. Particle $1$ is the strongest one of the whole system, while particle $N$ is the weakest one. Therefore,  matrix $B_N$ is a complete and symmetric matrix for egalitarian systems, while matrix $B_N$ is a lower-triangular matrix for dominance hierarchical system.

\subsection{Egalitarian \textit{versus} hierarchical systems}
Now we can give a formal definition of egalitarian and several kinds of hierarchical systems.

\subsubsection{Egalitarian system}
For each pair of individuals, if both of them will use each other's information with the same weight to decide the behavior at the next step, we say it is an \textit{egalitarian system}. An egalitarian system satisfies the following two rules.
\begin{enumerate}
  \item $c_{ij}=q$, $\forall i,j=1,\cdots,N$, $q$ is a positive constant;
  \item $b_{ij}=1$, $\forall i,j=1,\cdots,N$.
\end{enumerate}

\subsubsection{Contribution driven hierarchical system}
A contribution hierarchical system satisfies the following two rules.
\begin{enumerate}
  \item $c_{ij}$ follows some distribution for $i, j=1, \cdots, N$ (thus, not all of these $c_{ij}$ have the same positive value);
  \item $b_{ij}=1$, $\forall i,j=1,\cdots,N$,
\end{enumerate}

\subsubsection{Single-layer leader-follower hierarchical system (dominance driven hierarchical mechanism)}
A single-layer leader-follower hierarchical system satisfies the following two rules.
\begin{enumerate}
  \item $c_{ij}=q$, $\forall i,j=1,\cdots,N$, $q$ is a positive constant;
  \item $b_{ij}=1$, $i>j$, $i,j=1,\cdots,N$,
\end{enumerate}

\subsubsection{Double-layer leader-follower hierarchical system (contribution driven dominance hierarchical mechanism)}
In case the weights in the contribution matrix are not equal, we associate the system with the presence of dominance (the contribution of agents having a larger weight dominate over the contribution by those with a smaller weight). We consider such systems whose behavior is determined through a \textit{contribution driven} mechanism. Then, the particle with larger weight contribution is named as leader, while the other one is named as follower. A double-layer leader-follower hierarchical system satisfies the following two rules:
\begin{enumerate}
  \item $c_{ij}$ meets some distribution for $i, j=1, \cdots, N$ (but not all of these $c_{ij}$ have the same positive value);
  \item $b_{ij}=1$, $i>j$, $i,j=1,\cdots,N$,
\end{enumerate}

The above systems can be characterized by their intersection matrix $c_{ij}*b_{ij}$, $i,j=1,\cdots,N$. For an egalitarian system, it is a complete matrix with the same elements. For contribution hierarchical systems, it becomes a complete matrix with various elements. For single-layer leader-follower hierarchical system, it is a lower-triangular matrix with the same lower-triangular elements. For double-layer leader-follower hierarchical system, it becomes a lower-triangular matrix with varying lower-triangular elements.

Besides, in a more general model, $c_{ij}$ and $b_{ij}$ can be time-dependent. If $c_{ij}$ and $b_{ij}$ is not time-dependent that means everyone in the system has fixed a set of relationships. If $c_{ij}(t)$ and $b_{ij}(t)$ are time-dependent, then the relationships among these particles changes with time. Other different variants can thus be defined according to the contribution matrix $c_{ij}(t)$ and dominance matrix $b_{ij}(t)$. Here, we only discuss the case that when $c_{ij}$ and $b_{ij}$ are constant.

\section{Hierarchical model for flocking}

\begin{figure}[htbp]
\centering
\includegraphics[width=5in]{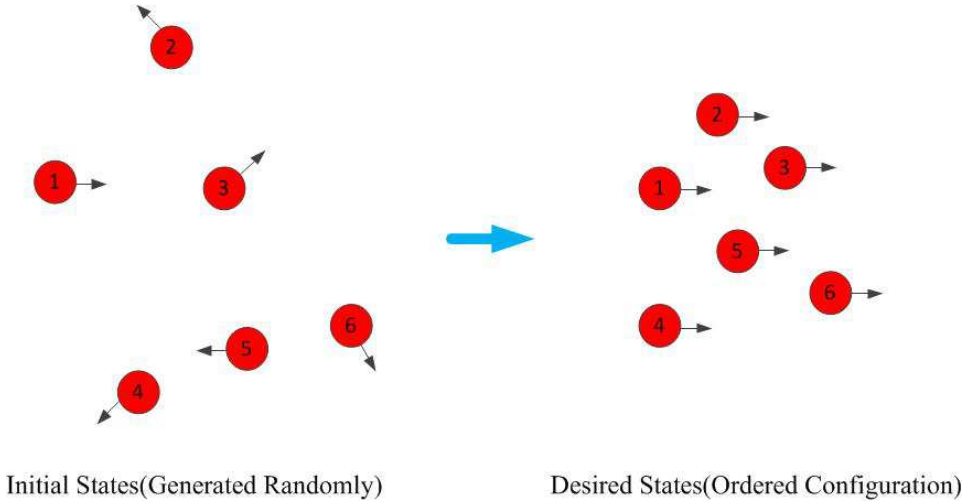}
\caption{initial configuration vs the ordered phase of collective motion.}
\label{cohesiveflocking}
\end{figure}

We consider in this paper $N$ particles moving continuously (off lattice) in a free area without any boundary limitation. As shown in Figure \ref{cohesiveflocking}, the position and direction of $N$ particles at the beginning are generated randomly, while over time these particles are expected to move coherently (ordered state). The figure is for the noiseless case.

Suppose that the time interval between two updates of the directions and positions is $\Delta t = 1$. This assumption can be made without loosing generality since $\Delta t$ occurs only in combination (being multiplied by it) with velocity terms.  At $t = 0$, $N$ particles were randomly distributed within an area of a given size and have the same absolute velocity $\upsilon$ as well as randomly distributed directions $\theta$. At each time step, the position of the $i$th particle is updated according to
\begin{equation}
\mathit{\mathbf{x}}_i(t+1) = \mathit{\mathbf{x}}_i(t) + \mathit{\mathbf{v}}_i(t)\Delta t
\end{equation}
In each time step, the velocity of a particle $\mathbf{v}_i(t+1)$ is updated according to the following equation
\begin{equation}
\label{velocity}
{\mathbf{v}}_i(t+1) = {\mathbf{v}}_i^{align}(t) + {\mathbf{v}}_i^{rep}(t) + {\mathbf{v}}_i^{adh}(t)
\end{equation}
where $\mathbf{v}_i^{align}(t)$ is the alignment term, $\mathbf{v}_i^{rep}(t)$ is the repulsion term, and $\mathbf{v}_i^{adh}(t)$ is the attraction term.

The alignment term was constructed to have an absolute value $\upsilon$ and a direction given by the angle $\theta^{align}_i(t+1)$. The angle was obtained from the expression
\begin{equation}
\theta^{align}_i(t+1) = <\theta_i(t)> + \Delta \theta(t)
\end{equation}
where $\Delta \theta(t)$ represents noise, which is a random number chosen with a uniform probability from the interval $[-\eta/2, \eta/2]$. $<\theta_i(t)>$ denotes the average direction of the velocities of neighbors of the given particle $i$. The average direction is given by the angle
\begin{equation}
<\theta_i(t)>= arctan\left(\frac{\sum\limits_{j=1}^N l_{ij}(t)sin(\theta_j(t))}{\sum\limits_{j=1}^N l_{ij}(t)cos(\theta_j(t))} \right).
\end{equation}
The matrix $\textit{L}_N(t)=[l_{ij}(t)]_{N\times N}$ describes the neighbor relationships of particles at time $t$, where
\begin{equation}
l_{ij}(t)= c_{ij} * b_{ij} * a_{ij}(t), \forall i, j = 1, \cdots, N.
\end{equation}
The definition of adjacency matrix $A_N(t)=[a_{ij}(t)]_{N \times N}$ is
\begin{equation}
\label{neighbor_matrix}
{a_{ij}(t)} = \left\{ {\begin{array}{*{20}{c}}{1, i=1, \cdots, N, \ \ j \in N_i(t)\ \ }\\
{0, otherwise \ \ \ \ \ \ \ \ \ \ \ \ \ \ \ \ \ \ \ \ }
\end{array}} \right.,
\end{equation}
where $N_i(t)=\{j|\|\mathbf{x}_{i}(t)-\mathbf{x}_{j}(t)\|<r$. Here, $r$ denotes the interaction radius. Using the above expressions the alignment term can be written as
\begin{equation}
\mathbf{v}_i^{align}(t+1) ={{\rm{c}}^{{\rm{align}}}}{\upsilon}{\mathbf{e}_i(t)}
\end{equation}
where ${{\rm{c}}^{{\rm{align}}}}$ is the coefficient of the alignment term. $\mathbf{e}_i(t)$ is a unit vector with direction angle $\theta_i^{align}(t)$.

The repulsion term exists only when the distance between any two particles is smaller than the repulsive radius $r_{rep}$. And the repulsion term is defined as
\begin{equation}
\mathbf{v}_i^{rep}(t+1) = {{\rm{c}}^{rep}}\sum\limits_{j=1}^N l_{ij}(t){\left( {\frac{{{r_{rep}} - \left\| {{\mathbf{x}_{ij}(t)}} \right\|}}{{{r_{rep}}}} \bullet \frac{{{\mathbf{x}_{ij}(t)}}}{{ \left\| {{\mathbf{x}_{ij}(t)}} \right\| }}} \right)},
\end{equation}
where $\left\| {{\mathbf{x}_{ij}(t)}} \right\| < r_{rep}$ and ${{\rm{c}}^{{\rm{r}}ep}}$ is the coefficient of the repulsion term.

The attraction term is only considered for the boundary particles \cite{Turner2014} of the whole system when the distance between two particles is between $r_{rep}$ and $r_{adh}$.
\begin{equation}
\mathbf{v}_i^{adh}(t+1) = {{\rm{c}}^{att}}\sum\limits_{j=1}^N l_{ij}(t){\left( {\frac{{{r_{rep}} - \left\| {{{\mathbf{x}}_{ij}(t)}} \right\|}}{{{r_{att}} - {r_{rep}}}} \bullet \frac{{{\mathbf{x}_{ij}}(t)}}{{\left\| {{\mathbf{x}_{ij}(t)}} \right\| }}} \right)},
\end{equation}
where ${r_{rep}} \le \left\| {{\mathbf{x}_{ij}(t)}} \right\| \le {r_{adh}}$ and ${{\rm{c}}^{{\rm{a}}dh}}$ is the coefficient of the attraction term. This term is introduced in order to prevent the flock spreading (or, in other words, "evaporating") due to perturbations.

\begin{figure}
\centering
\subfigure[$b_{ij}=q$, $\forall i,j$]{\includegraphics[width=2in]{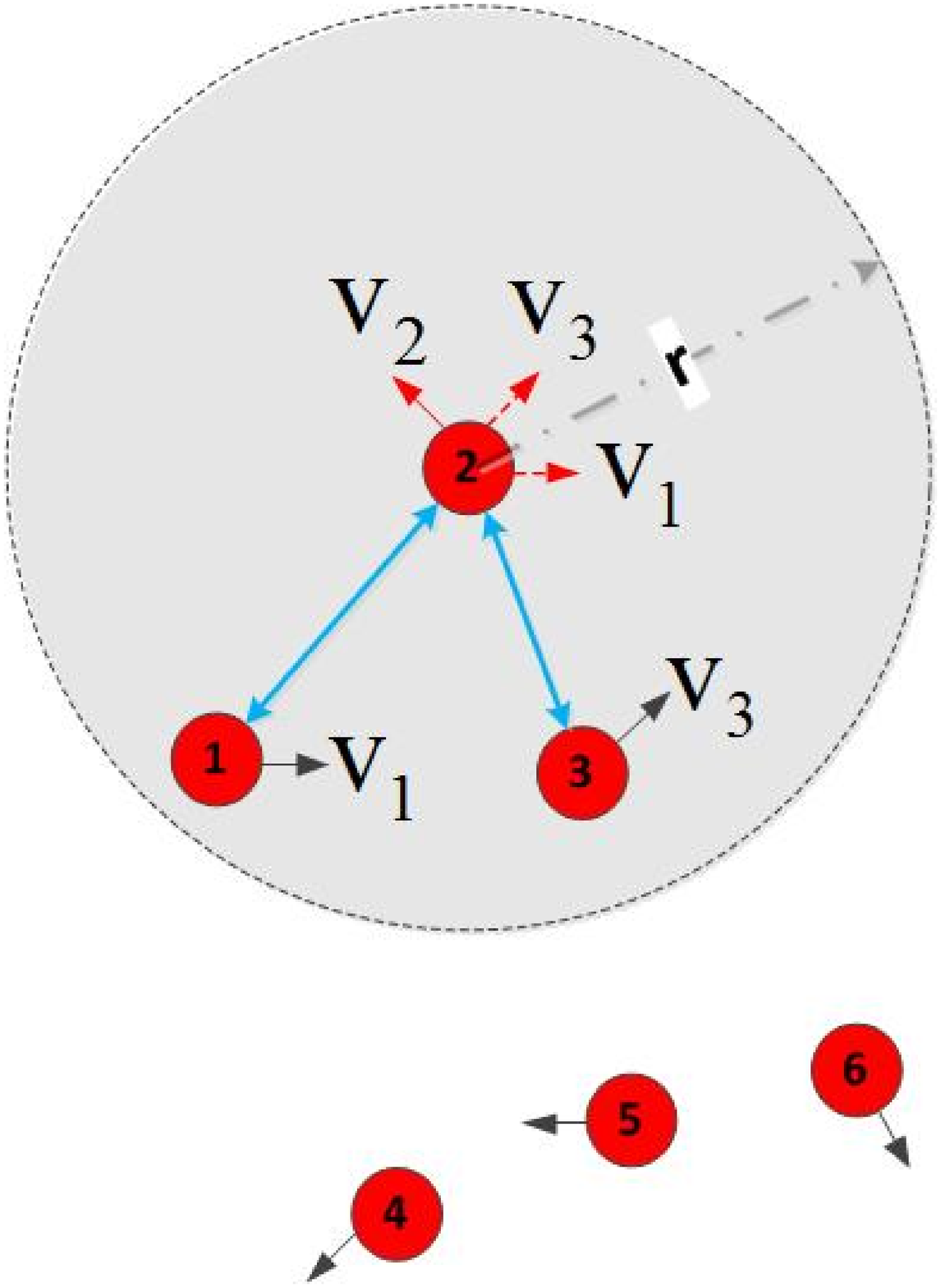}} \ \ \ \ \ \ \ \
\subfigure[$b_{ij}\neq q$, $\forall i,j$]{\includegraphics[width=2in]{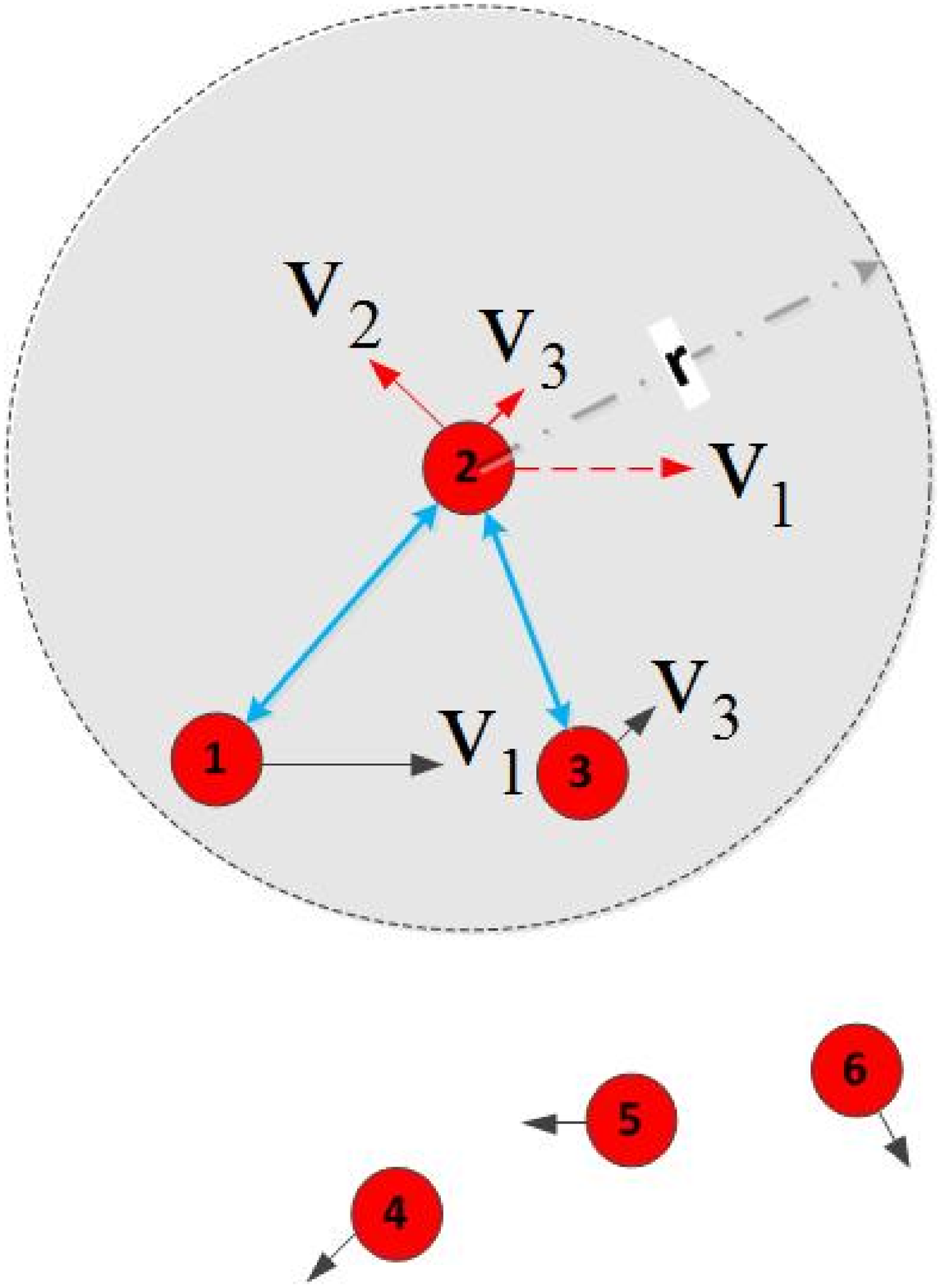}} \\
\subfigure[$b_{ij}=q$, $\forall i,j$ and $i>j$]{\includegraphics[width=2in]{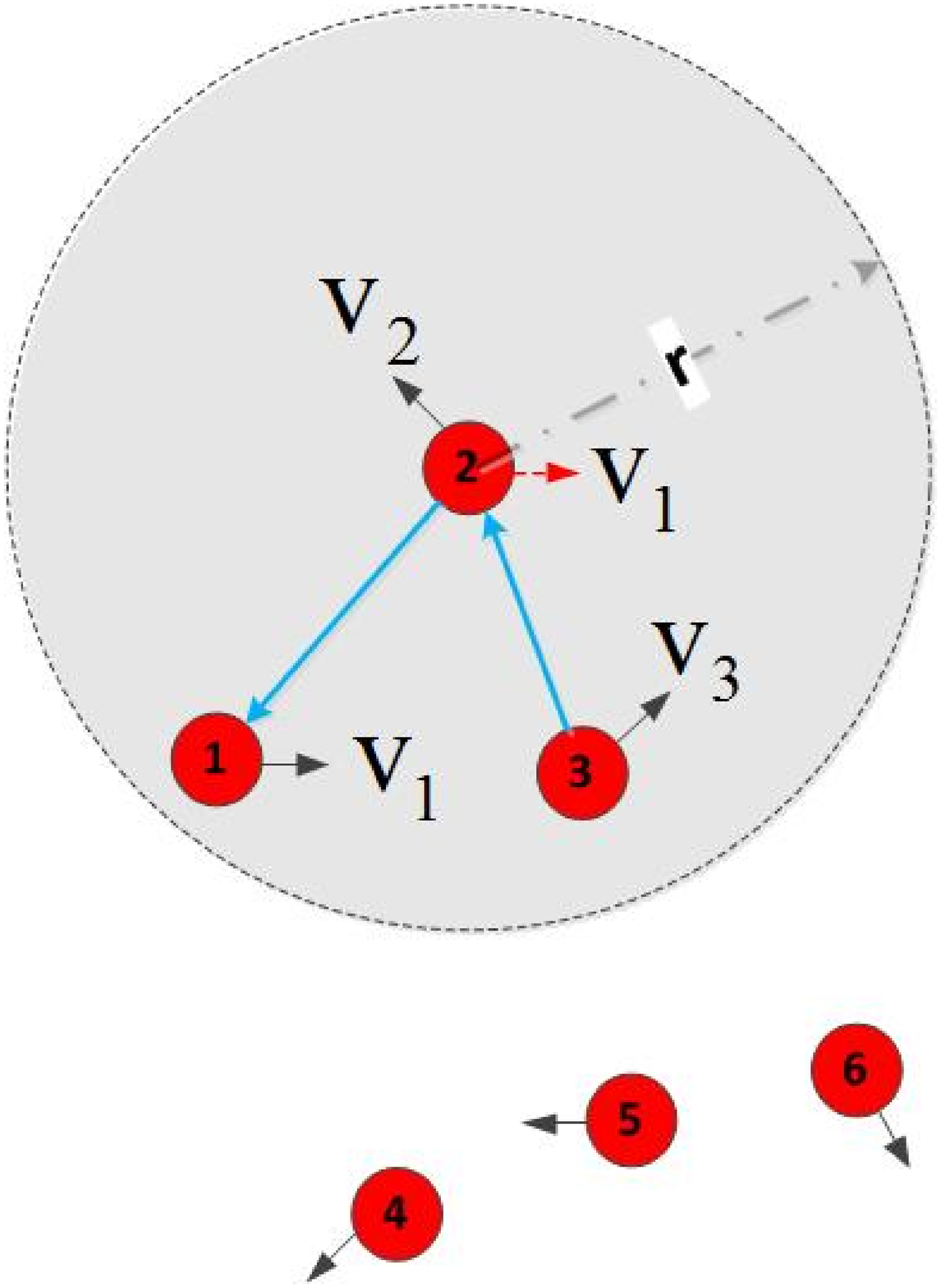}} \ \ \ \ \ \ \ \
\subfigure[$b_{ij}\neq q$, $\forall i,j$ and $i>j$]{\includegraphics[width=2in]{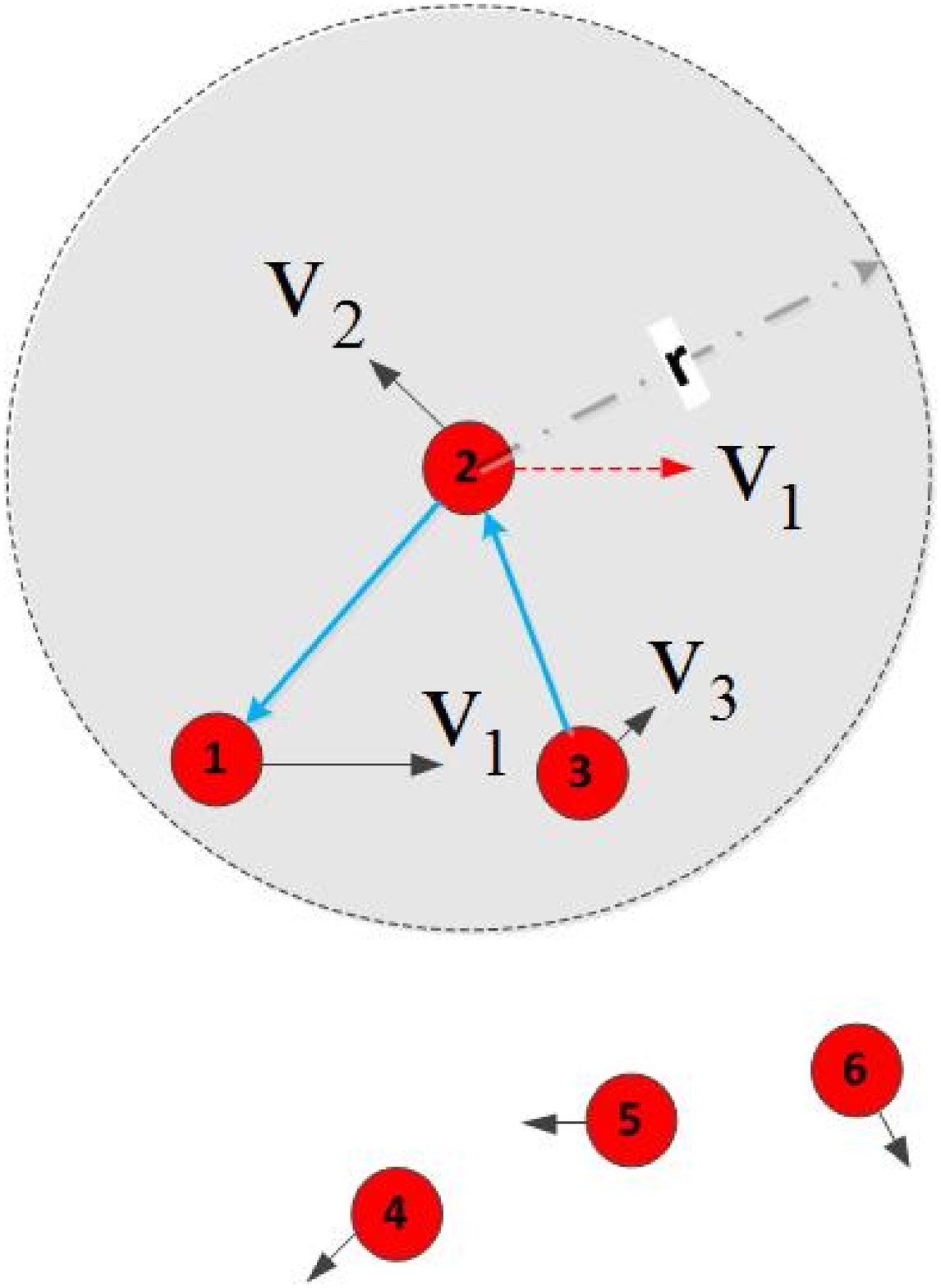}}
\caption{Neighbor matrix. (a)egalitarian system; (b)contribution driven hierarchical system; (c)single-layer leader-follower Hierarchical system; (d)double-layer hierarchical system.}
\label{neighbormatrix}
\end{figure}

Figure. \ref{neighbormatrix} shows the neighbor matrix of several variants of flocks mentioned in the last section. From Figure. \ref{neighbormatrix}, we can see more details on the difference among egalitarian system and other hierarchical system more clearly.

\section{Simulations and discussion}

The simulations were carried out in a free two-dimensional area. We considered groups of particles having various sizes ranging from $10$ particles to $1280$ particles. In order to keep the continuity and comparability of these simulation results of different group sizes, we chose the scale of the area for the random initial positions directly proportional to the scale of group size $N$, and generated the initial angle from the interval $(-\pi, \pi]$. In this simulation, we use $\upsilon=0.1$, $r_{rep}=0.5$, $r_{adh}=2.2$, and interaction radius $r=2.2$. We aimed at comparing the stability of egalitarian systems versus various hierarchical systems for varying levels of  external disturbance (noise) and for different group sizes.

The alignment item of egalitarian flock model is much like the self propelled particle model proposed by Vicsek et. al. in 1995 \cite{Vicsek1995}, where the definition of neighbors of particle $i$ includes itself and the contribution of the particles is the same. Thus, we call the egalitarian model as VEM (E for egalitarian, M for model). The contribution driven hierarchical model is called as CHM (C for contribution), while single-layer leader-follower Hierarchical model is named by SHM (S for single-layer leader follower).

According to the definition of the dominance matrix $B_N$, the neighbor matrix of a hierarchical system has zero-value diagonal elements, that is, we have $l_{ii}=0$, $\forall i=1,\cdots,N$ for all hierarchical systems. $l_{ii}=0$ means that the neighbor set of particle $i$ doesn't contain itself. In the following under hierarchy, we always mean hierarchical leader-follower kind of hierarchy. Therefore, we name the double leader-follower hierarchical flock model with zero-s along the diagonal ($l_{ii}=0$, $\forall i=1,\cdots,N$) as 'Double Hierarchical Model with Zero-s (DHMZ)', while when $l_{ii}\neq0$, $\forall i=1,\cdots,N$, we call it 'Double Hierarchical Model(DHM)'.

\subsection{VEM vs CHM/SHM}
\subsubsection{Order Parameter}
In this case we used the average normalized velocity is as the order parameter, defined as
\begin{equation}
\phi^{ave}=\frac{1}{T} \frac{1}{N} \int_0^T \|\sum_{i=1}^N {\mathbf{v}_i(t)}\| dt
\end{equation}
where $T=2000$ is the simulation time for each experiment.

The error bar is defined as:
\begin{equation}
\sigma = \frac{1}{\sqrt{n}}\bar{\phi}
\end{equation}
where $\bar \phi$ is the standard deviation of the order parameter values, and $n>0$ is the number of simulations for a given system size. The typical values of $n$ were chosen as follows:
\begin{equation}
{n} = \left\{ {\begin{array}{*{20}{c}}
{1000, if \ \ N=10 }\\
{1000, if \ \ N=20 }\\
{500, if \ \ N=40 }\\
{200, if \ \ N=80 }\\
{100,  if \ \ N=160 }
\end{array}} \right.
\end{equation}

\subsubsection{Results}
\begin{figure}[htbp]
\centering
\includegraphics[width=6.5in]{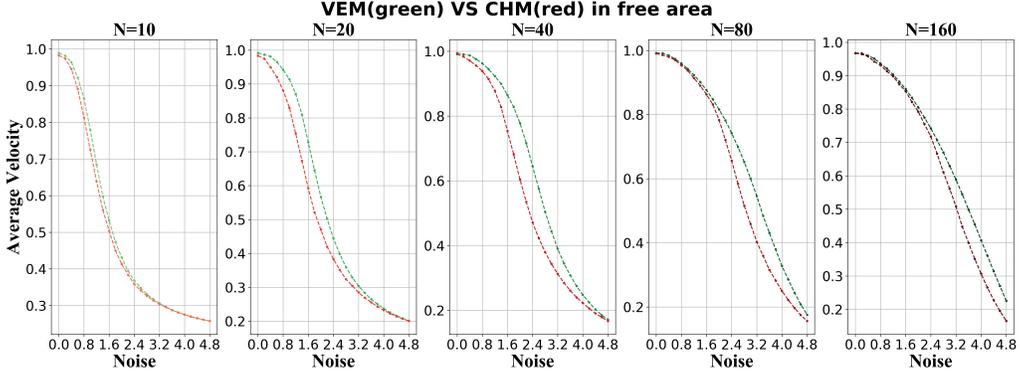}
\caption{Quantitative comparison on VEM and CHM ($l_{ii}\neq0$ for $i \in 1, \cdots, N$) in various of group sizes. $c_{ij}$ satisfies log-normal distribution (whose mean is $0$ and standard deviation is $1$) for the CHM system. Note that both Fig. 3 and 4. shows that the egalitarian system is more stable then the simple dominance-based hierarchical system.}
\label{VESvsCDS}
\end{figure}

\begin{figure}[htbp]
\centering
\includegraphics[width=6.5in]{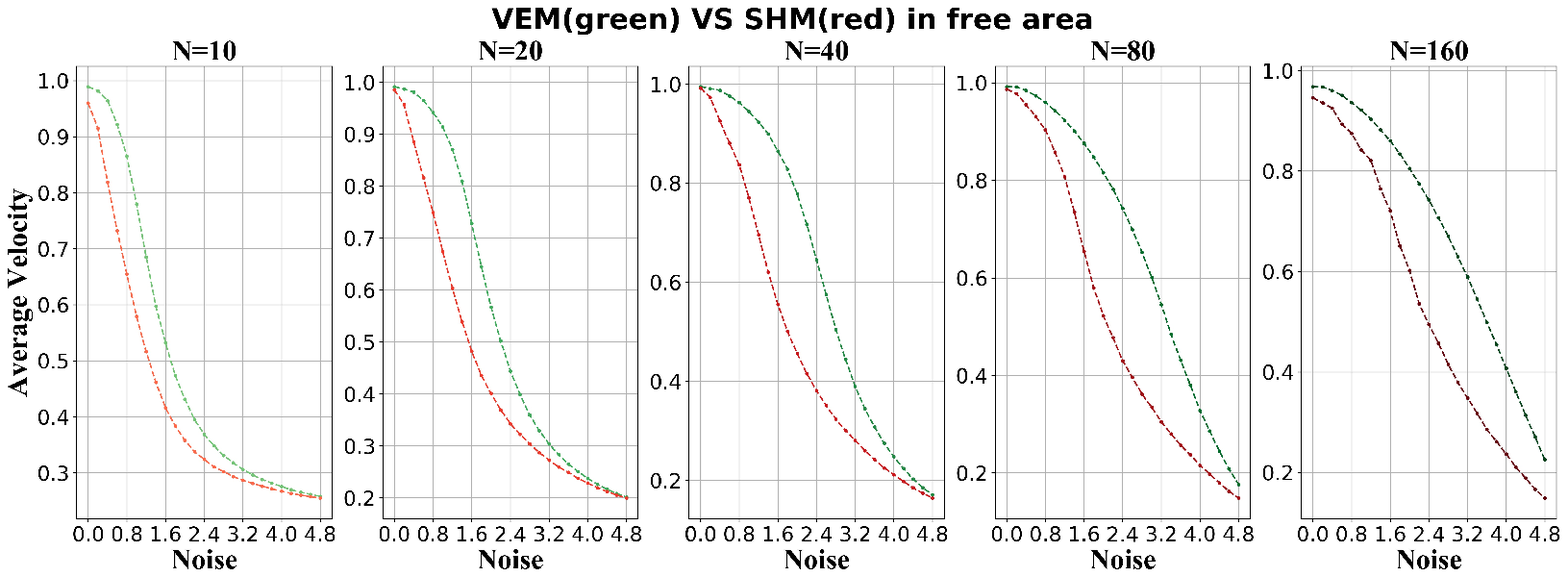}
\caption{Quantitative comparison of the VEM and SHM ($l_{ii}\neq0$ for $i \in 1, \cdots, N$) for various of group sizes. $c_{ij}$ satisfies log-normal distribution(whose mean is $0$ and standard deviation is $1$) for the SHM system.}
\label{VESvsSDS}
\end{figure}

We have compared the stability of VEM and some simple hierarchical flocking systems, such as CHM and SHM. According
to our results (see Figure. \ref{VESvsCDS} and Figure. \ref{VESvsSDS}), a simple dominance based system does not perform better than the much studied egalitarian flock.

In this paper we primarily report on our results concerning leader-follower systems. For more details about VEM versus CHM and SHM see hal.elte.hu/~vicsek/downloads/papers/Trieste-poster-JYN-TV-final.pdf

\subsection{VEM vs DHM/DHMZ}
\subsubsection{Order Parameter}
\begin{figure}[htbp]
\centering
\includegraphics[width=6in]{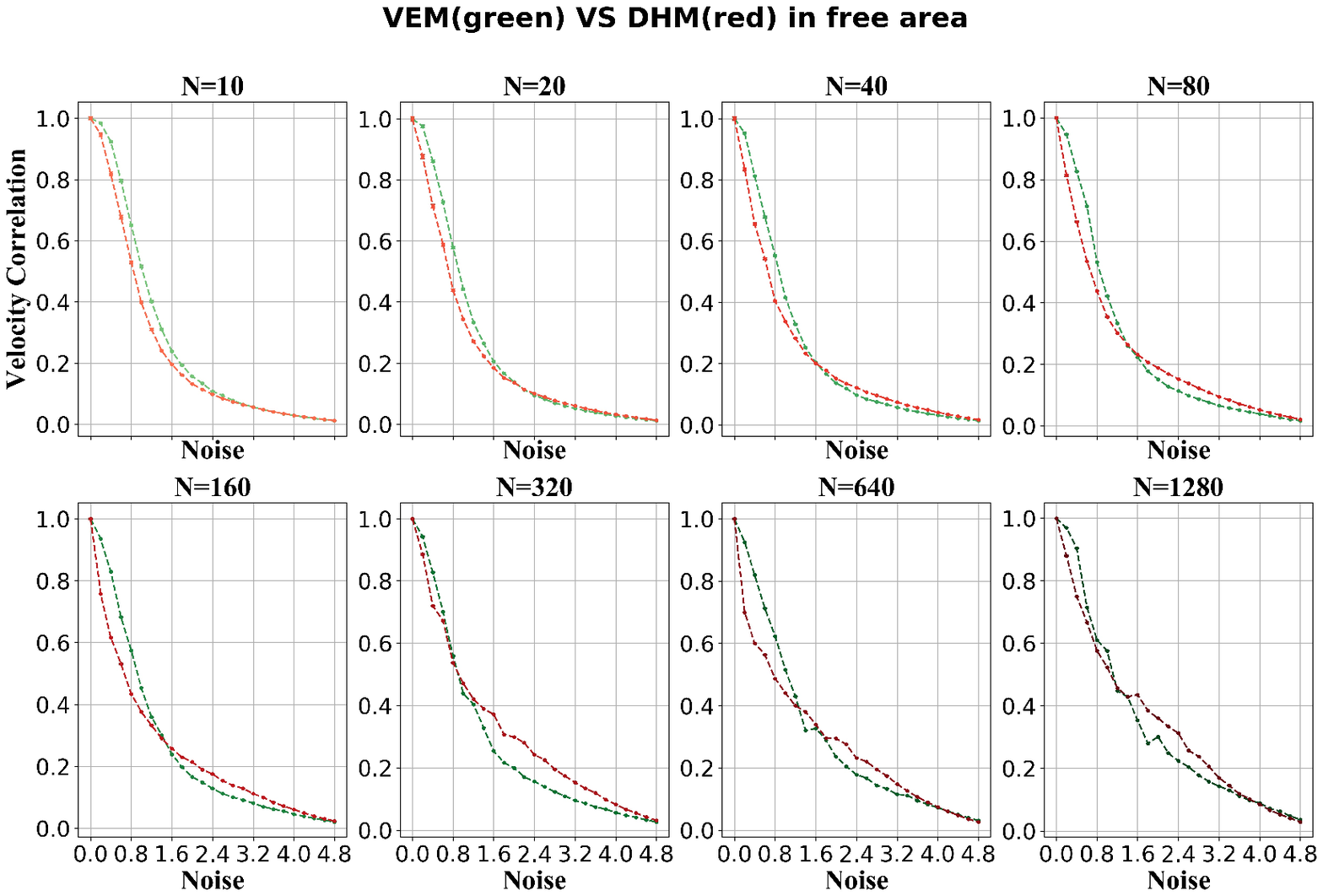}
\caption{quantitative comparison on VEM and DHM ($l_{ii}\neq0$ for $i \in 1, \cdots, N$) in various of group sizes. $c_{ij}$ satisfies log-normal distribution(whose mean is $0$ and standard deviation is $1$) for VDM system.}
\label{VESvsVDS}
\end{figure}
\begin{figure}[htbp]
\centering
\includegraphics[width=6in]{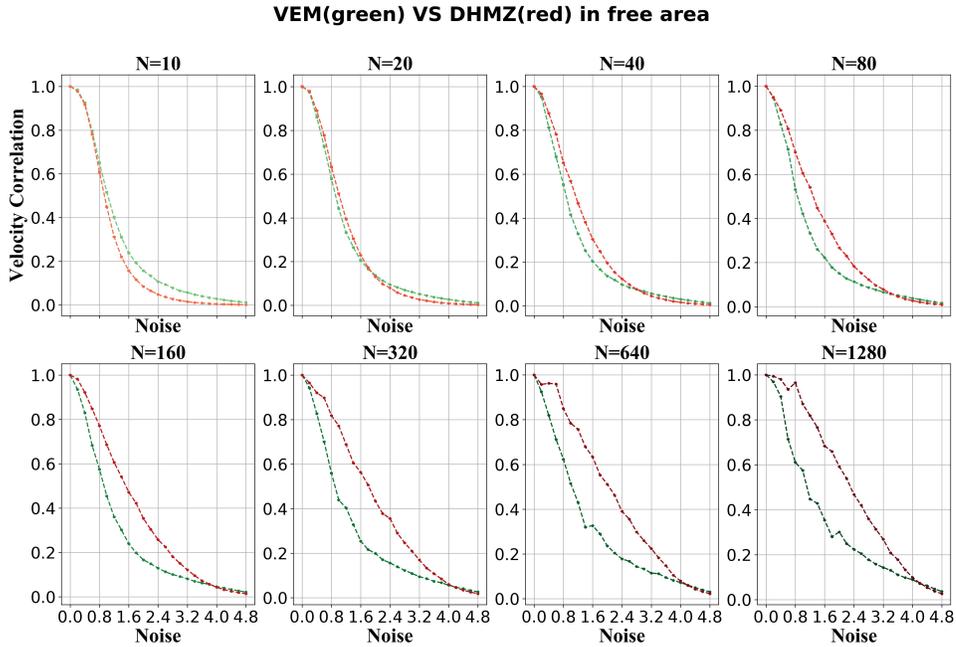}
\caption{quantitative comparison on VEM and DHMZ ($l_{ii}=0$ for $i \in 1, \cdots, N$) in various of group sizes. $c_{ij}$ satisfies log-normal distribution(whose mean is $0$ and standard deviation is $1$) for DHMZ system.}
\label{VESvsNDS}
\end{figure}
As mentioned above, we shall associate the stability with the tendency of a flock not to break into parts under external perturbations. In order to measure the stability of the particle group, we used the following velocity correlation as the order parameter
\begin{equation}
\phi^{corr}=\frac{1}{T}\frac{1}{N(N-1)}\int_0^T \sum_{i=1}^N \sum_{j \in \textit{N}_i(t)/i} \frac{\mathbf{v}_i(t)\mathbf{v}_j(t)}{\|\mathbf{v}_i(t)\|\|\mathbf{v}_j(t)\|}dt,
\end{equation}
where $j \in \textit{N}_i(t)/i$ means $j \in \textit{N}_i(t)$ and $j\neq i$. This expression indicates the stability of the flock under different conditions. We did not use the average velocity as the order parameter, because the average velocity cannot give a right stability description when the system is divided into two or more coherently moving subgroups.
%The typical values of $n$ are chosen as follows.
%\begin{equation}
%{n} = \left\{ {\begin{array}{*{20}{c}}
%{1000, if \ \ N=10 }\\
%{500, if \ \ N=20 }\\
%{250, if \ \ N=40 }\\
%{120, if \ \ N=80 }\\
%{60,  if \ \ N=160 }\\
%{10, if \ \ N=320 }\\
%{5, if \ \ N=640 }\\
%{2, if \ N=1280 }
%\end{array}} \right.
%\end{equation}

\subsubsection{Results}
We compare VEM with DHM and DHMZ separately. Figure. \ref{VESvsVDS} shows the comparison results of the VEM and DHM, while Figure. \ref{VESvsNDS} shows the comparison results of VEM and DHMZ.
\begin{figure}
\centering
\subfigure[$t=1s$]{\includegraphics[width=5.6in]{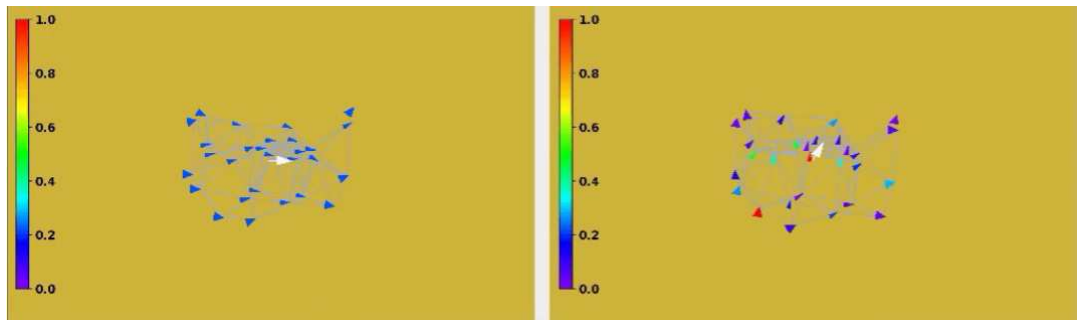}}
\subfigure[$t=27s$]{\includegraphics[width=5.6in]{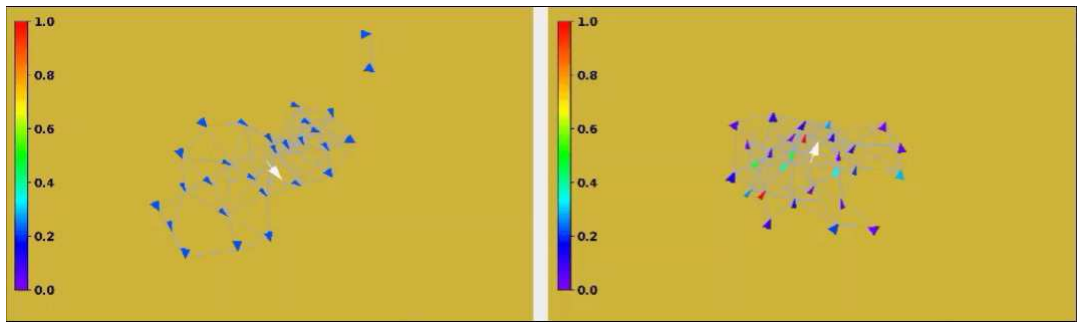}}
\subfigure[$t=53s$]{\includegraphics[width=5.6in]{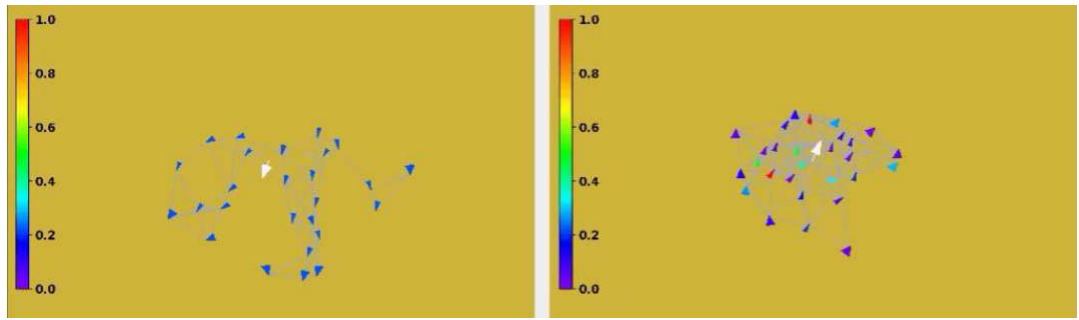}}
\subfigure[$t=56s$]{\includegraphics[width=5.6in]{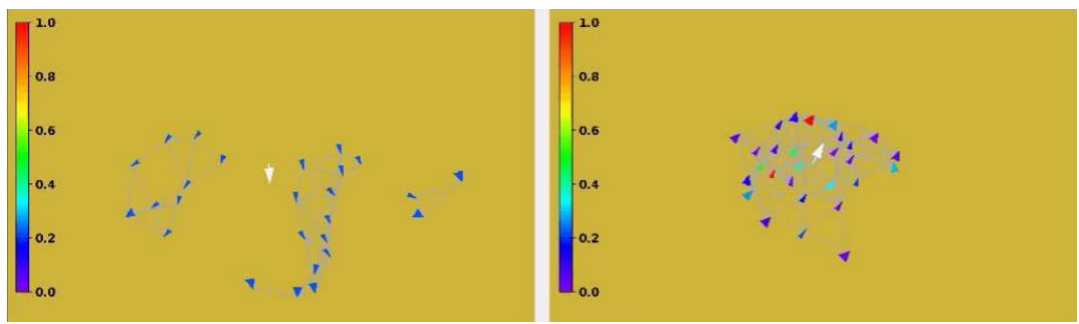}}
\caption{VEM and DHMZ for twenty-eight particles flocking. For each scenario, the left one is VEM, while the right one is DHMZ.(a) the scenario at time $t=1s$. (b) the scenario at first separate time $t=27s$. (c) the scenario at time $t=53s$. (d) the scenario at the second separate time $t=56s$.}
\label{28particles}
\end{figure}

According to Figure. \ref{VESvsVDS}, we can see that VEM is a little bit better (better meaning that more ordered for the same amount of noise) than the DHM for different group size under lower noise, while it seems that DHM is a little better than VEM under larger noise. However, the advantage is so small that can be ignored. At the same time, it seems that there exists no relevant difference among the simulation results for various group sizes.

Figure. \ref{VESvsNDS} demonstrates that the DHMZ hierarchical model is significantly more stable than an egalitarian model for flocking, and the advantage is increasingly obvious as the size of the system increases. However, the stability gap between an egalitarian flock and a hierarchical flock stops increasing from $320$ particles to $1280$ particles. That is to say, the difference in the performances between the egalitarian system and the hierarchical system is increasing with system size, but only up to a given flock size and is not changing as the group size reaches a certain threshold. This is quite along the intuitive picture which suggests that hierarchy may play a relevant role only up to - at most - a few hundred of flock members.

Let us consider twenty-eight particles, for example, Figure. \ref{28particles} shows some important scenarios during the flocking process (for a movie see Supplementary Material S1). The color bar indicates the weight of the contribution of the given particle. For example, the red particle is the strongest one, while the purple one is the weakest particle. The contribution values of these particles belonging to the DHMZ system satisfy log-normal distribution. That is, $c_{ij}$, $i,j=\{1,\cdots,N\}$, $i>j$  satisfy log-normal distribution, with mean value $0$ and standard deviation $1$. At the same time, the contribution value of each particle belonging to VEM system is equal, and the sum of the contribution values of VEM system is equal to the sum of the contribution values of the DHMZ system. The first picture in Figure. \ref{28particles} displays the initial state of all the particles at start. The second frame to the left depicts the structure of the VEM flock being less cohesive than the DHMZ (the right one). The third picture shows a key moment when VEM results in two separated groups while DHMZ results still in a single cluster. This can be taken as an important evidence to demonstrate that leader-follower hierarchical systems are more efficient regarding their stability against perturbations which the individual units are subject to.

%\section{Conclusions}

\section{Acknowledgments}
This work was partly supported by the following grants: U.S. Air Force, Office of Scientific Research grant no. FA9550-17-1-0037; H2020, EU FP7 RED-Alert No: 740688 and the (Hungarian) National Research, Development and Innovation Office grant No K128780.

%\section{Supplemental Material}
%Supplementary Video.

\section{References}

\bibliographystyle{unsrt}   % Include this if you use bibtex
\bibliography{sample.bib}            % and a bib file to produce the

\begin{thebibliography}{10}

\bibitem{Sumpter2010}
D~J~T Sumpter.
\newblock {\em Collective Animal Behavior}.
\newblock Princeton University Press, 2010.

\bibitem{Mate2010}
M~Nagy, Z~Akos, D~Biro, and T~Vicsek.
\newblock Hierarchical group dynamics in pigeon flocks.
\newblock {\em Nature}, 464(7290):890¡ª893, April 2010.

\bibitem{Vicsek2012}
T~Vicsek and A~Zafeiris.
\newblock Collective motion.
\newblock {\em Physics Reports}, 517(3):71--140, 2012.

\bibitem{Petit2010}
O~Petit and R~Bon.
\newblock Decision-making processes: the case of collective movements.
\newblock {\em Behavioural processes}, 84(3):635¡ª647, July 2010.

\bibitem{Kata2015}
K~Ozogany and T~Vicsek.
\newblock Modeling the emergence of modular leadership hierarchy during the
  collective motion of herds made of harems.
\newblock {\em J. Stat. Phys}, 158:628--646, 2015.

\bibitem{Seeley2010}
T~D Seeley.
\newblock {\em Honeybee Democracy}.
\newblock Princeton University Press, New Jersey, 2010.

\bibitem{Anna2013}
A~Zafeiris and T~Vicsek.
\newblock Group performance is maximized by hierarchical competence
  distribution.
\newblock {\em Nature communications}, 4, 2013.

\bibitem{Vasarhelyi2018}
G~Vasarhelyi, C~Viragh, G~Somorjai, T~Nepusz, A~E. Eiben, and T~Vicsek.
\newblock Optimized flocking of autonomous drones in confined environments.
\newblock {\em Science Robotics}, 3(20), 2018.

\bibitem{Floreano2015}
D~Floreano and R~J Wood.
\newblock Science, technology and the future of small autonomous drones.
\newblock {\em Nature}, 521:460¨C466, 2015.

\bibitem{Brambilla2013}
M~Brambilla, E~Ferrante, M~Birattari, and M~Dorigo.
\newblock Swarm robotics: A review from the swarm engineering perspective.
\newblock {\em Swarm Intelligence}, 7(1):1--41, 2013.

\bibitem{Anna2018}
A~Zafeiris and T~Vicsek.
\newblock {\em Why we live in hierarchies? A quantitative treatise}.
\newblock Springer, Berlin, 2018.

\bibitem{Couzin2005}
I~D Couzin, J~Krause, N~R Franks, and S~A Levin.
\newblock Effective leadership and decision making in animal groups on the
  move.
\newblock {\em Nature}, 433:513--516, 2005.

\bibitem{Sueur2008}
C~Sueur and O~Petit.
\newblock Organization of group members at departure is driven by social
  structure in macaca.
\newblock {\em Int J. Primatol.}, 29:1085--1098, 2008.

\bibitem{Sarova2010}
R~Sarova, M~Spinka, J~L~A Panama, and P~Simecek.
\newblock Graded leadership by dominant animals in a herd of female beef cattle
  on pasture.
\newblock {\em Animal Behaviour}, 79(5):1037 -- 1045, 2010.

\bibitem{Mate2013}
M~Nagy, G~Vasarhelyi, B~Pettit, I~Roberts-Mariani, T~Vicsek, and D~Biro.
\newblock Context-dependent hierarchies in pigeons.
\newblock {\em Proceedings of the National Academy of Sciences},
  110(32):13049--13054, 2013.

\bibitem{Shen2006}
J~Shen.
\newblock Cucker-{S}male flocking under hierarchical leadership.
\newblock {\em SIAM J. Appl. Math.}, 68(3):694--719, 2007/08.

\bibitem{Turner2014}
D~J~G Pearce and M~S Turner.
\newblock Density regulation in strictly metric-free swarms.
\newblock {\em New Journal of Physics}, 16, 07 2014.

\bibitem{Vicsek1995}
T~Vicsek, A~Czirok, E~Ben-Jacob, I~Cohen, and O~Shochet.
\newblock Novel type of phase transition in a system of self-driven particles.
\newblock {\em Physical Review Letters}, 75(6):1226, 1995.

\end{thebibliography}

\clearpage
\clearpage
\end{document}